\documentclass[preprint]{revtex4}
\usepackage{amsmath}
\usepackage{float}
\usepackage{graphics,epsfig}
\usepackage{graphicx}
\usepackage{dcolumn}
\usepackage{bm}
\usepackage{mathrsfs}
\usepackage{amssymb}
\usepackage{color}

\begin{document}
\baselineskip=0.8 cm
\title{{\bf Warm Inflation With A General Form Of The Dissipative Coefficient}}

\author{Yi Zhang}
\email{zhangyi@itp.ac.cn} \affiliation{Institute of Theoretical
Physics, Chinese Academy of Sciences, P.O.Box 2735, Beijing 100190,
China\\
Graduate University of Chinese Academy of Sciences, YuQuan Road 19A,
Beijing 100049, China}

\vspace*{0.2cm}
\begin{abstract}
\baselineskip=0.6 cm

We propose and investigate a general form of the dissipative
coefficient $\Gamma=C_{\phi}T^{m}/\phi^{m-1}$ in warm inflation. We
focus on discussing the strong dissipative processes
$r=\Gamma/3H\gg1$ in the thermal state of approximate equilibrium.
To this toy model, we give the slow-roll conditions, the amplitude
and the index of the power spectrum under the general form of
dissipative coefficient. Furthermore, the monomial potential and the
hybrid-like potential are analyzed specifically. We conclude that
the $m=0,3$ cases are worthy further investigation especially.
\end{abstract}

\maketitle

\section{Introduction}\label{sec1}
Since the introduction of inflation
scenario~\cite{Guth:1980zm,Linde:1981mu}, inflation has been
recognized as an essential stage in the early universe. In most
inflation models, the scalar field potentials are responsible for
accelerating the universe, and the primeval radiation would be
red-shifted. The consequence is that the universe becomes cold and
we can safely neglect the effect from the temperature. Therefore,
the process called ``reheating'', is needed to make the universe hot
again after inflation. From the observational aspect, up to now, we
still lack strong evidences  to support the point that the universe
is cold during inflation. From the theoretical aspect, there also
exist some controversy about the temperature of the universe during
inflation~\cite{Moss:1985wn}. A new type of inflation model called
``warm inflation'', in which the effects from radiation and
temperature are considered during inflation, was proposed by Berera
and Fang~\cite{Berera:1995ie,Berera:1995wh}.  In warm inflation
scenario, the acceleration of the universe is still driven by the
potential energy density of the scalar field which is called
inflaton, however, because of the interactions between inflaton and
other fields, the radiation can not be red-shifted heavily and the
universe is hot during inflation. The recent development about warm
inflation is nicely reviewed in~\cite{Berera:2008ar}.

Strictly speaking, in warm inflation scenario, the acceleration of
the universe should be driven by the free energy density of the
thermal system rather than the ordinary potential energy density of
the inflaton $\phi$. The Hubble parameter $H$ should  be expressed
as
 \begin{equation} \label{H}
  H^{2}=\frac{1}{3m_{pl}^{2}}(\frac{\dot{\phi}^{2}}{2}+V(\phi,T)+T s),
 \end{equation}
where  $m_{pl}$ is the planck mass, $T$ and $s$ are the temperature
and the entropy density of the thermal system respectively. Here, we
must point out that the potential $V(\phi,T)$ in Eq. (\ref{H}) is
the effective finite temperature potential, which is not only the
function of $\phi$ but also the function of temperature $T$.
However, under some special backgrounds, such as SUSY, the finite
temperature corrections to the potential $V(\phi,T)$ can be
suppressed dramatically, so under these backgrounds we can
approximatively use the zero temperature potential $V(\phi)$.

After considering the dissipative interaction, the equation of
motion of the inflaton becomes
\begin{equation} \label{me}
\ddot{\phi}+(3H+\Gamma)\dot{\phi}+V_{\phi}=0,
\end{equation}
where $\Gamma $ is the dissipative coefficient, which is related to
the microscopic physics of the interactions, and the subscript
$\phi$ means the partial derivative with respect to $\phi$. The
energy conservation equation of the radiation gives
 \begin{equation} \label{rc}
 \dot{\rho_{r}}+4H\rho_{r}=\Gamma \dot{\phi}^{2},
 \end{equation}
 where the right hand side term $\Gamma
\dot{\phi}^{2}$ is the source of the radiation. The hydrodynamic
description of the radiation gives the relationship between its
energy density and the universal temperature~\cite{early}
  \begin{equation}
  \label{rT} \rho_{r}=C_{r}T^{4},
 \end{equation}
where $C_{r}=\pi^{2}g^{*}/30$, $g^{*}$ is the effective number of
the light degrees of freedom in the universe. In the Minimal
Supersymmetric Standard Model (MSSM), $g^{*}=228.75$ and
$C_{r}=75.18$, however, if the effective temperature during
inflation is lower than the typical temperature of MSSM, $g^{*}$
could be smaller than order ${\cal O}(10)$, and $C_{r}$ is of order
${\cal O}(1)$. In convenience, we define a dimensionless parameter
$r=\Gamma/3H$, then $r>1$ corresponds to a strong dissipative
process, $r<1$ corresponds to a weak dissipative process, and
$r\ll1$ returns to usual ordinary inflation case. For simplicity, we
restrict our discussions under the very strong dissipative case
$r\gg1$.

To quantify the dissipative coefficient, we list some results in
different models. And the dissipative coefficient mainly depends on
the temperature and the amplitude of the inflaton field. With
different decaying mechanism,  different forms of $\Gamma$  should
be utilized. The forms of dissipative coefficient can be divided
into two classes especially in the low temperature, based on the
SUSY background or the non-SUSY background. Firstly, we give out the
results in the  SUSY background. Following Ref. \cite{Moss:2006gt},
the interactions are given in  the super potential
    \begin{equation}
    \label{in1} W=g\Phi X^{2}+hXY^{2},
  \end{equation}
where $g$ and $h$ are coupling constants of order ${\cal O}(1)$,
$\Phi$, $X$, $Y$ are super fields,  and the scalar components of the
super fields are $\phi$, $\chi$, $y$ respectively. During inflation,
the field $y$ and its fermionic partner $\tilde{y}$ remain massless;
while the mediating field $\chi$  gets its mass from the interaction
with the inflaton $\phi$,  $m_{\chi}\approx g\phi$, and its bosonic
partner $\tilde{\chi}$ also has a mass of $m_{\tilde{\chi}}\approx
g\phi$.  When the dissipation works with an intermediate
boson($\phi\rightarrow 2y$ and $\phi\rightarrow 2\tilde{y}$), at the
high temperature where $h^{-1}m_{\chi}>T\gg m_{\chi}$, $\Gamma
\approx 0.691g^{2}T/h^{2}$, and at the low temperature where $T\ll
m_{\chi}$, $\Gamma
\approx0.04g^{2}h^{4}(g\phi/m_{\chi})^{4}(T^{3}/m_{\chi}^{2})$. When
the dissipation works with an intermediate fermion ($\phi\rightarrow
y\tilde{y}$) at the high temperature where $h^{-1}m_{\chi}>T\gg
m_{\chi}$, $\Gamma \approx 0.97g^{2}T/h^{2}$; at the low temperature
where $T\ll m_{\chi}$, $\Gamma \approx g^{-2}h^{4}T^{5}/\phi^{4}$
which can be ignored when the dissipation working with an
intermediate boson happen at the same time. Furthermore when the
dissipation works with the exponential decay propagator, it gives
out $\Gamma\approx
h^{2}g^{3}\phi(2+g\phi/m_{\chi}+g^{3}\phi^{3}/m_{\chi}^{3})/16\pi^{2}$
in the zero-temperature limit
\cite{Hall:2004ab,Berera:2003kg,Berera:2002sp}. Secondly, in the
non-supersymmetry background, the dissipative coefficient $\Gamma$
has such  a form $\Gamma \approx C_{\phi}\phi^{2}/T$
~\cite{Yokoyama:1998ju, Berera:1998gx}, where $C_{\phi}$ is ${\cal
O}(1)$, when there is only one decaying field and one decaying
channel.

Considering all the different forms of dissipative coefficients
above, we propose such a general form
 \begin{equation}
\label{gf}\Gamma=C_{\phi}\frac{T^{m}}{\phi^{m-1}},
 \end{equation}
where $C_{\phi}$ is connected to the dissipative microscopic
dynamics, and except in
 the case of $m=0$  where $C_{\phi}$ is an order ${\cal O}(10^{-2})$
 parameter, we can take $C_{\phi}$ to be of order ${\cal
O}(1)$, if there was only one field and one channel to decay. With
the number of the decaying fields and their channels increasing,
$C_{\phi}$ becomes very large, and a natural upper bound could be
$C_{\phi}<10^{4}$ \cite{BasteroGil:2006vr}. From the discussions
above, we can see that, when $m=-1$, $\Gamma=C_{\phi}\phi^{2}/T$,
the form corresponds to the dissipative coefficient in the non-SUSY
case, when $m=0$, $\Gamma=C_{\phi}\phi$, it corresponds to the SUSY
case with an exponentially decaying propagator, when $m=1$,
$\Gamma=C_{\phi}T$, it corresponds to the high temperature SUSY
case, when $m=3$, $\Gamma=C_{\phi}T^{3}/\phi^{2}$, it corresponds to
the low temperature SUSY case.

In principle, when the inflaton interacts with other fields, whether
the dissipative productions are in an equilibrium process should be
determined by the detailed microphysics. However, in our toy model,
we just assume that the dissipative coefficient takes the assumed
general form  that the dissipative process is close to thermal
equilibrium, without any calculations in the microphysics aspect.
For convenience, we only focus on the parameter regime $|m|<4$. And
especially, we will pay attention to the cosmological results, focus
on the potential takes the monomial form and the hybrid-like form
respectively when $m=-1,0,1,3$.

Warm inflation  has two main advantages.
  $\Gamma\dot{\phi}^{2}$ in Eq. (\ref{me}) as an
additional term compared with ordinary inflation, will make the
slow-roll conditions not so restrictive. And for the existence of
temperature during inflation, if the thermal fluctuations are much
larger than the quantum fluctuations, density perturbations will be
reproduced by the thermal fluctuations, rather than the usual
quantum fluctuations.

Here we mostly discuss how  warm inflation  work with a general form
of  dissipative coefficient $\Gamma=C_{\phi}T^{m}/\phi^{m-1}$. We
will start with the slow-roll conditions that make sure the
inflation phase enough long in Sec. \ref{sec2}. And the analytic
results for the power spectrum, in particular its amplitude and its
index, will be given in Sec. \ref{sec3}. For a matter of
application, we will focus on two specific potentials, the monomial
potential in Sec. \ref{sec4} and the hybrid-like potential in Sec.
\ref{sec5}. At last, concise
 summary will be provided in Sec. \ref{sec6}.


\section{Slow roll conditions}
\label{sec2}

In the framework of warm inflation, the scalar potential is still
needed to drive the acceleration phase as in the ordinary inflation.
Then we also need the slow-roll conditions
\cite{Moss:2008yb,Moss:2007cv,Hall:2003zp} which technically
neglecting the $\dot{\phi}^{2}$ and $Ts$  in Eq. (\ref{H}),
$\ddot{\phi}$ in Eq. (\ref{me}), and $\dot{\rho_{r}}$ in Eq.
(\ref{rc}) will ensure sufficient e-folding number, stability of the
system. Due to the existence of finite temperature and the new
dissipative coefficient, there are three additional slow-roll
parameters $\beta, b, c$~\cite{Moss:2008yb}. The slow-roll
parameters can be divided into those related to the field $\phi$ and
those related to the temperature $T$.

 The slow-roll parameters related to  $\phi$ are:
 \begin{equation} \label{sl5}
\epsilon=\frac{m_{pl}^{2}}{2}(\frac{V_{\phi}}{V})^{2},
\eta=m_{pl}^{2}\frac{V_{\phi \phi}}{V},
\beta=m_{pl}^{2}\frac{V_{\phi}\Gamma_{\phi}}{V\Gamma},
\end{equation}
 and the warm inflation scenario requires
the slow-roll conditions $\epsilon, |\eta|,|\beta| \ll 1+r$. In the
strong dissipative regime $r\gg1$, the conditions can be simplified
to
 \begin{equation}
 \label{sl1}
 \epsilon, |\eta|, |\beta|\ll r.
 \end{equation}
It shows, just as we mentioned before, that the slow-roll conditions
for $\epsilon$ and $\eta$ are much looser than those in ordinary
inflation.

The other two additional parameters $b$ and $c$  related to the
temperature can be expressed as:
\begin{equation}
 \label{sl2}
 b=\frac{TV_{\phi T}}{V_{\phi}}, c=\frac{T\Gamma_{T}}{\Gamma}
 \end{equation}
  \begin{equation} \label{c}
 0<b\ll \frac{r}{1+r}, |c|<4,
 \end{equation}
where the subscript $T$ means partial derivative with respect to
$T$. The parameter $b$, which shows the extent that the scalar
potential is affected by the interactions, can satisfy Eq. (\ref{c})
if a mechanism for suppressing the thermal corrections is known in
the dissipative processes. As we mentioned in the introduction, such
dissipative processes, in which the temperature corrections to
effective potential are suppressed, can be realized in SUSY
background~\cite{BasteroGil:2006vr,BuenoSanchez:2008nc,Berera:2002sp}.
However, when $m=-1$, there are still some problems for the
realization of warm inflation under the SUSY
background~\cite{Yokoyama:1998ju}. In this paper, we will not
discuss the specific dissipative mechanism, and we just assume that
there is certain mechanism which makes $b$ appropriate for the slow
roll condition for simplicity.

With slow-roll conditions,  we can ignore the gravitational friction
term
 $3H\dot{\phi}$ and the second order derivative
term $\ddot{\phi}$ in Eq. (\ref{me}), then
 \begin{equation}\label{phi}
\dot{\phi}=-\frac{V_{\phi}}{\Gamma}.
 \end{equation}
Considering Eqs. (\ref{rT}) and (\ref{phi}),  the temperature of the
universe during inflation is
 \begin{equation} \label{T}
T=(\frac{V_{\phi}^{2}\phi^{m-1}}{4HC_{\phi}C_{r}})^{1/(4+m)} .
 \end{equation}

The parameter $\beta$ gives a new constraint on the evolution of
$\Gamma$. Combined Eq. (\ref{sl5}) with Eq. (\ref{T}), $\beta$ can
be rewritten as
  \begin{equation}
  \label{beta}
 \beta=\frac{2m}{4+m}\eta-\frac{m}{4+m}\epsilon-
 \frac{4(m-1)}{4+m}\frac{m_{pl}}{\phi}\sqrt{2\epsilon}.
 \end{equation}
If the term $m_{pl}/\phi$ depends on $\epsilon$ or
 $\eta$ as we will  discuss in Sec. \ref{sec4} and Sec. \ref{sec5},
 $\beta$ can be expressed in terms of $\epsilon$ and $\eta$.

In our case, the parameter $c$ which gives another new constraint on
the evolution of $\Gamma$, can give a strong constraint on the
potential. By virtue of Eq. (\ref{T}), Eq. (\ref{sl2}) can be
rewritten as
\begin{equation}
c=m+\frac{(4+m)(1-m)}{m-1+(2\phi V_{\phi\phi})/V_{\phi}-(\phi
V_{\phi})/(2V)}.
 \end{equation}

In addition, in warm inflation, the potential of the scalar field
still dominates the universe during inflation, so $\rho_{r}/V<1$ is
the basic requirement, furthermore the workable temperature regions
are crucial too. Following, we will discuss the value of
$\rho_{r}/V$ and the value of $T/\phi$ in specific models.


\section{Power spectrum}
\label{sec3}

For a scalar field, if $T>H$, the amplitude of thermal fluctuations
will be larger than that of quantum fluctuations
\cite{Berera:1995wh}, then the main contribution to the energy
 density perturbations comes from the thermal fluctuations.
The amplitude of the spectrum in the strong dissipative regime
($r\gg1$) is
\cite{Berera:1999ws,Hall:2003zp}
\begin{equation}\label{ps1}
 P_{R}^{1/2}\simeq |\frac{H}{\dot{\phi}}|(\frac{\pi
r}{4})^{1/4}\sqrt{T H}=|\frac{3H^{3}}{V_{\phi}}|(\frac{\pi
r}{4})^{\frac{1}{4}}(1+r)\sqrt{\frac{T}{H}}.
 \end{equation}
The observations tell us that the order of  the amplitude of power
spectrum should be  $P_{R}^{1/2}\simeq 5\times10^{-5}$. Compared
with ordinary inflation, the amplitude of power spectrum in warm
inflation does not only depend on the Hubble parameter $H$ and the
potential $V$ but also depend on the dissipative coefficient
$\Gamma$ which is contained in $r$ and the temperature $T$. In Eq.
(\ref{ps1}), large $T$ and $r$ are not favored by a small amplitude
of the spectrum, and we will discuss this problem in the concrete
models in detail.

We can calculate the  index of the spectral in the strong dissipative regime by using the definition
\begin{equation}
n_{s}-1=2\frac{d\ln P_{R}^{1/2}}{d\ln
k}=2\frac{\dot{\phi}}{H}\frac{d\ln P_{R}}{d\phi}
=\frac{-2V_{\phi}}{3H^{2}(1+r)}\frac{d \ln P_{R}^{1/2}}{d\phi},
 \end{equation}
where $k$ is the wavenumber, and that combined with Eqs. (\ref{gf}),
(\ref{T}) and (\ref{ps1}), we can get
 \begin{eqnarray}
 \label{index}
\nonumber  n_{s}-1&=&
\frac{-1}{1+r}[\frac{-6-2m+mC}{4+m}\eta+\frac{(-2-m)C+19+5m}{4+m}\epsilon\\&
&+\frac{(-m+1)(-1+2C)}{4+m}\frac{m_{pl}}{\phi}\sqrt{2\epsilon}],
\end{eqnarray}
where
\begin{equation}
 C=\frac{(8+m)(r+1)+4r(3+m)}{(4+m)(r+1)-2m}.
 \end{equation}
As $r\gg 1$, $C\approx5$. Further simplifying Eq. (\ref{index}), we
can get
\begin{equation}
  n_{s}-1 \approx \frac{-1}{1+r}[\frac{-6+3m}{4+m}\eta+\frac{9}{4+m}\epsilon+\frac{9(1-m)}{4+m}\frac{m_{pl}}{\phi}\sqrt{2\epsilon}].
 \end{equation}
And if we rewrite the above equation, according to Eq. (\ref{beta}) we can get
\begin{equation}
  n_{s}-1 \approx \frac{1}{1+r}(\frac{2}{3}\eta-\frac{9}{4}\epsilon-\frac{9}{4}\beta)
\end{equation}
which agrees with the previous results in \cite{Moss:2008yb,BasteroGil:2006vr,BuenoSanchez:2008nc,BasteroGil:2004tg}.
The WMAP$5$ data  require $n_{s}\simeq 1$ for a near flat power
spectrum, more specifically, $0.949<n_{s}<0.977$ if the running of
the spectral is forbidden, and $0.976<n_s<1.085$ with a running of
the spectral $-0.065<dn_{s}/d\ln k<-0.009$ \cite{Komatsu:2008hk}.
The running of the spectral can be expressed as
 \begin{eqnarray}
 \label{run}
\nonumber n_{s}'&=& \frac{dn_s}{d ln k}=\frac{1-n_{s}}{1+r}r'
-\frac{1}{1+r}[\frac{-6+3m}{4+m}\eta'+\frac{9(1-m)}{4+m}(\frac{m_{pl}}{\phi}\sqrt{2\epsilon})'\\
& &+\frac{9}{4+m}\epsilon' +\frac{m
D}{4+m}\eta+\frac{2(1-m)D}{4+m}\frac{m_{pl}}{\phi}\sqrt{2\epsilon}-\frac{(2+m)D\epsilon}{4+m}],
 \end{eqnarray}
where
 \begin{eqnarray} \nonumber
D=C' \approx\frac{3+m}{4+m}\frac{5r'}{r},
 \end{eqnarray}
\begin{equation}
 \eta'=-\frac{\epsilon}{1+r}(\xi-2\eta),
  \end{equation}
\begin{equation}
\epsilon'= -\frac{2\epsilon}{1+r}(\eta-2\epsilon),
 \end{equation}
 \begin{equation}
(\frac{m_{pl}}{\phi}\sqrt{2\epsilon})'=\frac{m_{pl}}{\phi}\frac{\sqrt{2\epsilon}}{1+r}(\frac{m_{pl}}{\phi}\sqrt{2\epsilon}+2\epsilon-\eta),
 \end{equation}
\begin{equation}
r'=\frac{[(4+2m)\epsilon+(4m-4)\sqrt{2\epsilon}m_{pl}/\phi-2m\eta]r}{(4+m)r+4-m},
 \end{equation}
where  $'= d / d\ln k$, $\xi= 2m_{pl}^{2}V_{\phi \phi
\phi}/V_{\phi}$. When $dn_s/d\ln k<0$, the running of the spectrum
is negative, and relatively when $dn_s/d\ln k>0$, the running of the
spectrum is positive. From Eq. (\ref{index}) if both $\epsilon\sim
\eta$ and $m_{pl}\sqrt{2\epsilon}/\phi \sim \eta$, $n_s-1$ will be
of order ${\cal O}(\eta/r)$, and $dn_s/d\ln k$ will be of order
${\cal O}(\eta^{2}/r^{2})$. And if $\eta$  is of order ${\cal
O}(1)$, as $r\gg1$, we will probably get a near flat spectrum.  The
$\eta$ problem which can not get a flat power spectrum when $\eta$
at order ${\cal O}(1)$ in the ordinary inflation is alleviated. When
the slow-roll conditions in Eq. (\ref{sl1}) are satisfied, the
spectrum is probably flat and the running  could be small. As we
see, the expression of the running is very complicated. In concrete
potentials, we will use the exact form of $n_s$  to carry on our
calculations.

As in Ref. \cite{Moss:2008yb}, the power spectrum of tensor
perturbations is just the same as in the ordinary inflation
$P_{T}=H^{2}/(2\pi^{2}m_{pl}^{2})$, so the tensor-to-scalar
amplitude ratio is
\begin{equation}
\frac{P_{T}}{P_{R}}=\frac{2}{\pi^{5/2}}\frac{\epsilon}{(1+r)^{2}r^{1/2}}\frac{H}{T}.
 \end{equation}
If $\epsilon/r<1$, as $T/H>1$ and $r\gg 1$, the tensor-to-scalar
amplitude ratio will be  much smaller  than that in the ordinary
inflation where $P_{T}/P_{R}=12.4\epsilon$ \cite{Lyth:1998xn}. This
result could probably meet the observations ($P_{T}/P_{R}<0.43$
without a running of the spectrum, and $P_{T}/P_{R}<0.58$ with a
running of the spectrum \cite{Komatsu:2008hk}).

In the next two sections, we will apply the above equations to the
monomial potential and the hybrid-like potential respectively. To
get a realizable  warm inflation, we will discuss about four
dominant conditions: the slow-roll conditions $\epsilon, |\eta|,
|\beta|\ll r$ and $|c|<4$; $T/H>1$ which makes the density
 spectrum  originated from the thermal fluctuations;  the observable
amplitude of the density perturbations  $P_{R}^{1/2}\approx
5\times10^{-5}$ which should be satisfied;  the workable temperature
regime which should be suitable as well.


\section{Monomial potential}
\label{sec4}

The form of monomial potential reads
\begin{equation}
 \label{mv}
 V(\phi)=V_{0}(\frac{\phi}{m_{pl}})^{n},
 \end{equation}
where $V_0$ and $n$ are the two free parameters of the theory. As we
know, when $n=2$, it is the well known chaotic inflation, and when
$n=4$, it represents the self interactions of the inflaton. In this
paper, we only consider the cases $n\geq2$ in this kind of
potential. In ordinary inflation the workable regime of this kind of
models is $\phi>m_{pl}$ in which the causal problem would
arise~\cite{linde}. We put this potential in warm inflation
background to check whether the inflation can work when we require
$\phi\leq m_{pl}$ and $V(\phi)<m_{pl}^{4}$ simultaneously.

From Eq. (\ref{me}), we can get that the potential deceases from its
initial value $V_i$ as
\begin{equation}
  \label{mv0}\frac{V}{V_i}= (1-\frac{4N_{e}\rho_{ri}}{V_i}
  \frac{12+m n-2n-4m}{n(4+m)})^{n(4+m)/(12+m n-2n-4m)},
 \end{equation}
 where $N_{e}$ is the e-folding number with the definition
 $N_{e}=\int^{t_{end}}_{t}H dt$, the subscript $i$ means the value at the initial time.
 If $V/V_i\ll 1$,  the potential could not dominate the energy density of the universe during inflation.
  To make the model workable, we must have $V/V_i\sim{\cal
O}(10^{-1})$.

Since the dissipative coefficient is related to the temperature, it
is necessary to calculate the value of $T/\phi$ for discussing the
workable temperature regime. Eqs. (\ref{me}), (\ref{T}) and
(\ref{mv}) yield
\begin{equation} \label{mTphi}
\frac{T}{\phi}=[(\frac{\sqrt{3}n^{2}}{4C_{\phi}C_{r}})^{2}
(\frac{V_0}{m_{pl}^{4}})^{3}(\frac{\phi_i}
{m_{pl}})^{(3n-8-2m)}(\frac{V}{V_i})^{(3n-14)/n}]^{1/2(4+m)}.
 \end{equation}
The evolution of $T/\phi$  depends on the value of $n$ and $m$
sensitively. Assuming $\phi_i\approx m_{pl}$ and $V_i\ll
m_{pl}^{4}$, we can get $T/\phi<1$, and the warm inflation works in
the low temperature regime. And as Eq. (\ref{mTphi}) shows, if
$\phi_i\ll m_{pl}$, warm inflation may take place in the high
temperature regime too when $n< (8+2m)/3$. For convenience, we only
consider the $m=-1,0,1,3$ cases in the following.

\subsection{Slow Roll Conditions in Monomial Potential}
\label{sec4a} Considering  the slow-roll parameters in the  monomial
potential, $\epsilon$
 and
$\beta$ can be expressed in terms of $\eta$
\begin{equation}
 \label{mee}
 \epsilon=\frac{n}{2(n-1)}\eta,
 \end{equation}
 \begin{equation}
 \beta=\frac{3mn-12m+8}{2(n-1)(4+m)}\eta.
 \end{equation}
Then as $n\geq2$, $|m|<4$, we can get $\epsilon \leq \eta$, $\beta$
is of order ${\cal O}(\eta)$. And we only need to check whether
$|\eta/r|<1$ while considering the slow-roll parameters related to
$\phi$. In this potential  $\eta>0$, $|\eta/r|=\eta/r$, we will use
$\eta/r<1$ below. In strong dissipation, from Eqs. (\ref{T}) and
(\ref{mv}), we can get
\begin{equation} \label{meta}
\eta=n(n-1)(\frac{V_i}{V})^{2/n}(\frac{m_{pl}}{\phi_i})^{2},
 \end{equation}

\begin{eqnarray} \label{mr}
 r=[\frac{ C_{\phi}^{4}}{9}(\frac{n^{2}}{4C_{r}})^{m}
(\frac{m_{pl}}{\phi_i})^{-4+6m}
(\frac{V_i}{m_{pl}^{4}})^{m-2}(\frac{V_i}{V})^{(-mn+2n-4+6m)/n}]^{1/(4+m)},
 \end{eqnarray}
\begin{eqnarray}
 \label{metar}
 \frac{\eta}{r}&=&
4\frac{n-1}{n}[\frac{9n^{8}C_{r}^{m}}{16C_{\phi}^{4}}
(\frac{V_{0}}{m_{pl}^{4}})^{2-m}
(\frac{\phi_i^{n}}{m_{pl}^{n}}\frac{V}{V_i}
)^{(2n-mn-12+4m)/n}]^{1/(4+m)}.
 \end{eqnarray}
By virtue of Eqs. (\ref{mv0}) and (\ref{metar}), $\eta/r<1$ is
equivalent to
\begin{equation}
4^{(m+2)/4}\sqrt{3n^{2}}C_{r}^{m/4}(N_{e}\frac{|12+m
n-2n-4m|}{n(4+m)}+1-\frac{1}{n})^{(m+4)/4}(\frac{m_{pl}}{\phi_i})^{3-m}(\frac{V_i}{m_{pl}^{4}}
)^{(2-m)/4}<C_{\phi},
 \end{equation}
 so $C_{\phi}$ is bounded below.

 From Eqs. (\ref{phi}) and (\ref{metar}), we can get $\rho_{r}/V= [n /(4n-4)](\eta/r) $, which
 means that
if the slow-roll conditions are satisfied, $\rho_r/V<1$ will be
satisfied automatically.

Then we check the slow-roll parameter $c$. In monomial potential
\begin{equation}
 |c|=|m+\frac{(1-m)(4+m)}{m-3+3n/2}|<4,
 \end{equation}
which gives a strong constraint on the model parameter $n$ directly.
Table \ref{tab1} denotes the suitable $n$ which is required by
appropriate $c$ when $m=-1,0,1,3$. When $m=-1,0$ the chaotic case
($n=2$) is excluded by the constraint on $c$. And the $n=4$ case can
satisfy the requirements of $c$ when $m=-1,0,1,3$.
 \begin{table}[ht]
\caption{The suitable range of $n$ for $c$, the value of $n_{s}-1$,
 the proper range of $n$ for the red spectrum and the blue
spectrum  in the monomial potential.}
\begin{center}
  \begin{tabular}{l@{\hspace{2mm}}l@{\hspace{2mm}}l@{\hspace{2mm}}l@{\hspace{2mm}}l@{\hspace{2mm}}l@{\hspace{2mm}}cc}
    \hline
&$m$&$|c|<4$&$n_{s}-1$&red spectrum  &blue spectrum  \\
\hline
 &$-1$&$n>52/15$ &$\frac{-\eta}{1+r}\frac{12-3n}{2(n-1)}$ &$2\leq n< 4$&$n> 4$ \\
 &$0$&$n>8/3$ &$\frac{-\eta}{1+r}\frac{21-3n}{8(n-1)}$&$2\leq n< 7$&$n> 7$ \\
 &$1$&$n\geq2$ &$\frac{-\eta}{1+r}\frac{6+3n}{10(n-1)}$&$2\leq n$&none \\
 &$3$&$n\geq2$&$\frac{-\eta}{1+r}\frac{-24+9n}{14(n-1)}$&$n> \frac{8}{3}$&$2\leq n< \frac{8}{3}$ \\
\hline
 \hline
  \end{tabular}
\end{center}\label{tab1}
\end{table}

\subsection{Power Spectrum in Monomial Potential}
\label{sec4b}
To study the energy density perturbations which come
from the thermal fluctuations, we list the analytic form of $T/H$:
\begin{eqnarray} \label{mTH}
\frac{T}{H}=[3^{5+m}(\frac{n^{2}}{4C_{\phi}C_{r}})^{2}(\frac{\phi_i^{n}}{m_{pl}^{n}}\frac{V}{V_i})^{(6-2m-mn-n)/n}(\frac{m_{pl}^{4}}{V_{0}})^{1+m)}]^{1/2(4+m)},
 \end{eqnarray}
and the amplitude of the power spectrum as well
\begin{eqnarray} \label{mps}
 P_{R}^{1/2}=(\frac{\pi}{4n^{4}})^{1/4}[3^{-m+13}C_{\phi}^{18}(\frac{n^{2}}{4C_{r}})
 ^{5m+2}
 (\frac{V_{0}}{m_{pl}^{4}})^{6m-3}
(\frac{\phi_i^{n}}{m_{pl}^{n}}\frac{V}{V_i})^{(42-28m+6mn-3n)/n}]^{1/4(4+m)}.
 \end{eqnarray}

 \begin{table}[ht]
\caption{The value of $dn_{s}/d\ln k$, the suitable range of $n$ for
the negative running and the positive running of the spectra in
monomial potential.}
\begin{center}
  \begin{tabular}{l@{\hspace{2mm}}l@{\hspace{2mm}}l@{\hspace{2mm}}l@{\hspace{2mm}}l@{\hspace{2mm}}l@{\hspace{2mm}}cc}
    \hline
&$m$&$dn_{s}/d\ln k$&negative running&positive running  \\
\hline
 &$-1$ &$\frac{\eta^{2}}{r^{2}}\frac{(3n-16)(4-n)}{(n-1)^{2}}$&$2\leq n< 4$, $n>\frac{16}{3}$&$4<n<\frac{16}{3}$ \\
 &$0$ &$\frac{\eta^{2}}{r^{2}}\frac{(n-6)(21-3n)}{8(n-1)^{2}}$&$2\leq n< 6$, $n>7$&$6<n<7$ \\
 &$1$ &$\frac{\eta^{2}}{r^{2}}\frac{(n-8)(6+3n)}{25(n-1)^{2}}$&$2\leq n< 8$&$n>8$ \\
 &$3$&$\frac{\eta^{2}}{r^{2}}\frac{-n(5n-6)}{49(n-1)^{2}}$&$n> \frac{8}{3}$&$2\leq n<\frac{8}{3}$ \\
\hline
 \hline
  \end{tabular}
\end{center}\label{tab2}
\end{table}

Combining Eq. (\ref{index}) with Eq. (\ref{mee}), the potential
gives $nm_{pl}\sqrt{2\epsilon}/\phi=\eta$, and we can yield the
index of the spectrum
\begin{equation} \label{mns}
 n_{s}-1=
\frac{-\eta_{H}}{1+r}\frac{6mn+21-15m-3n}{2(m+4)(n-1)},
 \end{equation}
which leads to the running of the spectrum as {\begin{equation}
\frac{dn_{s}}{d \ln
k}=\frac{\eta^{2}}{r^{2}}\frac{(6mn+21-15m-3n)(2n-m
n-12+4m)}{(4+m)^{2}(n-1)^{2}}.
 \end{equation}
$n-1=0$ is  not a real singularity for the two equations above
because in Eq. (\ref{metar}) the $\eta/r$ contain the $n-1$ term too
that can eliminate the singularity. We can see that these formulas
depend on $m$ sensitively. Considering that $\eta/r$ is positive,
the scopes of $n$ for both the red spectrum and the blue spectrum
are given in Table \ref{tab1}, and the proper regions of $n$ for
both the negative running and the positive running of the spectrum
are all listed in Table \ref{tab2}. When $n=2$, the spectrum is red
with a negative running in the $m=-1,0,1$ cases, and blue with a
positive running in the $m=3$ case. When $n=4$, the spectrum is
exactly flat in the $m=-1$ case, blue in the $m=0,1$ cases, and red
in the $m=3$ case, with no running in the $m=-1$ case and  a
negative running in the $m=0,1,3$ cases.

 We discuss the spectrum in the $m=-1,0,1,3$ in detail below. When $m=-1$,
\begin{equation} \label{mTH1}
\frac{T}{H}=(\frac{9n^{2}}{4C_{\phi}C_{r}})^{1/3}
(\frac{\phi_{i}}{m_{pl}}\frac{V}{V_{i}})^{4/3}.
 \end{equation}
Assume that
 $C_{r}$ and $C_{\phi}$ at order ${\cal O}(10)$, as
 $\phi_i/ m_{pl}\leq 1$,  and with $\phi$ rolling down, the potential energy of $\phi$ decreases,
$V/V_i<1$,  so the terms on the right hand side of  Eq. (\ref{mTH1})
 are not bigger than $1$, then $T/H>1$ is impossible.
 Therefore
warm inflation can not produce an observable power spectrum caused
by the thermal fluctuations when $m=-1$ and $T/H<1$ in the monomial
potential.

When $m=0$, $T/H>1$ yields
\begin{equation} \label{m0etar}
\frac{V_0}{m_{pl}^{4}}<3^{5}(\frac{n^{2}}{4C_{\phi}C_r})^{2}
(\frac{\phi_i^{n}}{m_{pl}^{n}}\frac{V}{V_i})^{(6-n)/n}.
 \end{equation}
 Combining the above expression with Eq. (\ref{mps}), we can get
\begin{equation}
 \frac{\phi_i}{m_{pl}}<\left [0.49n^{-14}C_{\phi}^
 {15}C_{r}^{5}(\frac{V}{V_i})^{3/2n}\right ]^{3/16}P_{R}^{1/3}.
 \end{equation}
 If we assume
the terms in the square brackets are at order  ${\cal O}(1)$, we
will get $\phi_i/m_{pl}<10^{-3}$ which is much different from the
ordinary inflation case where $\phi_i/m_{pl}>1$, so in this model we
can avoid the causal problem. Further considering the parameter
$V_{0}$, taking $n=2$ for example, we can get
$V_{0}^{\frac{1}{4}}<10^{15}GeV$ from Eq. (\ref{m0etar}). In
addition, $\eta/r<1$ requires the amplitude of the spectrum
 $P_{R}^{1/2}>6(n-1)^{9/8}(n/C_{r})^{1/8}(\phi_i/m_{pl})^{(3n-6)/8}(V_{0}/m_{pl}^{4})^{3/8}$.
 If the parameter value is as small as we discussed before, it is
 easy to satisfy this relation, and the $m=0$ case can work in the low temperature regime
\cite{Moss:2006gt}. Combined with Eq. (\ref{mTphi}),  Eq.
(\ref{m0etar}) gives
\begin{equation}
\frac{T}{\phi}<\frac{9n^{2}}{4C_{\phi}C_r}(\frac{V}{V_i})^{1/2n}(\frac{\phi_i}{m_{pl}})^{5/4}.
 \end{equation}
If $\phi_i /m_{pl}\leq1$,  $T/ \phi$ could be smaller than $1$, the model can also work in
the low temperature regime, so that the warm inflation can be
realized. And we notice that the above discussions do not give an obvious constraint
 to the parameter $C_{\phi}$, whether $C_{\phi}$ is larger or smaller than $10^{4}$,
  this scenario can both work.

When $m=1$, $T/H>1$ gives
\begin{equation}
\frac{V_{0}}{m_{pl}^{4}}<\frac{27n^{2}}{4C_{\phi}C_{r}}(\frac{m_{pl}^{n}}{\phi_i^{n}}\frac{V_i}{V})^{(n-2)/n},
 \end{equation}
and $\eta/r>1$ gives
\begin{equation}
\frac{V_{0}}{m_{pl}^{4}}<576^{-1}n^{-3}(n-1)^{-5}
C_{\phi}^{4}C_{r}^{-1}(\frac{m_{pl}^{n}}{\phi_i^{n}}\frac{V_i}{V})^{(n-8)/n}.
 \end{equation}
If the two conditions are satisfied, they give  the amplitude of the
spectrum two upper bounds. Because the observation constraints are
more strict $P_{R}^{1/2}\approx5\times10^{-5}$,  in fact the above
constraints are too loose to be effective.

The analyzed amplitude of power spectrum in this model is:
\begin{equation}
 P_{R}^{1/2}=
[1.1n^{-3/10}C_{r}^{-7/20}C_{\phi}^{9/10}
(\frac{\phi_i^{n}}{m_{pl}^{n}}\frac{V}{V_{i}})^{(14+3n)/20n}](\frac{V_{0}}{m_{pl}^{4}})^{3/20}.
 \end{equation}
We assume $\phi_i\approx m_{pl}$ and the terms in the square
brackets are of order  ${\cal O}(1)$, the observable amplitude of
the power spectrum requires
$V_{0}^{\frac{1}{4}}\approx8\times10^{10}GeV$. And as mentioned in
Eq. (\ref{mTphi}), under that assumption  the model favors a low
temperature case when $m=1$. But the dissipative processes need a
high temperature regime when $m=1$. Therefore, if we want to realize
warm inflation, we need a smaller $\phi_i$, not only to get a
suitable power spectrum, but also to get a relative high temperature
in the $m=1$ case, but that  is only useful when $n<10/3$ as Eq.
(\ref{mTphi})  shown. And whether $C_{\phi}$ is larger or smaller
than $10^{4}$ is not a problem for this case too.

When $m=3$,  a very detailed discussion has been done  in Ref.
\cite{BasteroGil:2006vr}.
 In our calculation, $T/H>1$ yields
\begin{equation}
\frac{V_{0}}{m_{pl}^{4}}<4.5(\frac{n^{2}}{C_{\phi}C_r})^{1/2}(\frac{V_i}{V})^{1/n}\frac{m_{pl}}{\phi_i},
 \end{equation}
which is easy to satisfy if we choose a low value of  $V_{0}$. And
$\eta/r<1$ requires
\begin{equation}
\frac{V_{0}}{m_{pl}^{4}}>9216n(n-1)
^{7}C_r^{3}C_{\phi}^{-4}\frac{V_i}{V}(\frac{m_{pl}}{\phi_i})^{n}.
 \end{equation}
Then we use the above inequality to replace $V_{0}/m_{pl}^{4}$ term
in $P_{R}^{1/2}$, and then we can get
\begin{equation}
P_{R}^{1/2}>1.64n^{7/4}(n-1)^{15/4}C_{\phi}^
{-3/2}C_{r}(\frac{m_{pl}}{\phi_i})^{3/2}(\frac{V}{V_i})^{-3/2n}.
 \end{equation}
In the above expression, assume $\phi_i/m_{pl}\approx1$ and $C_{r}$
is at order ${\cal O}(10)$, $C_{\phi}$ must be  at least at order
${\cal O}(10^{4})$ to get  low enough value of $P_{R}^{1/2}$, just
as Ref. \cite{BasteroGil:2006vr} mentioned. If $\phi_i$ is smaller
than $m_{pl}$, $C_{\phi}$ will be larger than ${\cal O}(10^{4})$
which is too large to be natural. Therefore, we exclude the $m=3$
case.

The above discussions are based on two assumptions, one is that $\phi$
is smaller than $m_{pl}$, which can avoid the causal problem in
ordinary inflation, the other is that the remained parameters also
satisfy the requirements from the particle physics including
$V_{0}<m_{pl}^{4}$ and a natural choice of the number of the inflaton fields
$C_{\phi}<10^{4}$. We find some regimes for monomial potential
the  warm inflation can work. The $m=0$ case
is workable. When $m=1$, the regime $2\leq n<10/3$ can work as well.
And in the $m=-1$ case, the warm inflation can not be realized
because $T/H<1$. In the $m=3$ case, the constraints from $C_{\phi}$
excludes the warm inflation.

\section{Hybrid-like potential}
\label{sec5}
 We now turn to the hybrid-like potential in  warm
inflation which is particularly motivated from particle physics.
During the inflation phase the potential of the hybrid-like
inflation is effectively described by a single field,
 \begin{equation} V(\phi)=V_{0}(1+\alpha
\ln \frac{\phi}{M}), n=0,
 \end{equation}
\begin{equation}
V(\phi)=V_{0}(1+(\frac{\phi}{M})^{n}),n>0,
 \end{equation}
 $\alpha$ is the
coupling constant, in the Yukawa coupling or the gauge coupling
which is of order ${\cal O}(1)$, and $M$ is a model parameter which
depends on the exact physical mechanism and we just treat as an
undetermined parameter. When $n=0$, the potential is derived from
the one-loop correction. And in the $n=2$ case, the potential
presents the standard hybrid-like model. However, there are some
constraints on the parameters from the point of view of particle
physics. Firstly, since
 the physics above the Planck scale is unknown, the scale of the
potential should be $V(\phi)<m_{pl}^{4}$. Secondly, when $n\geq4$,
the parameters must satisfy the condition
$V_{0}/(M^{n}m_{pl}^{4-n})<1$, which means that the coupling
constant should not be  too large to make the field $\phi$
ill-defined \cite{Lyth:1998xn}. On the other hand
$V_{0}/(M^{n}m_{pl}^{4-n})$ should not be too small to make the
coupling constant fine-tuning. In this paper, we assume $\phi \ll
M$, $\phi\leq m_{pl}$ simultaneously and $V_{0}$ is the main
contribution to $H^{2}$ where $H^{2}\approx V_{0}/3m_{pl}^{2}$. As
inflation continues, the field $\phi$ decreases during inflation.

We define a dimensionless  parameter $a$ for convenience,
\begin{equation}
a^{2}=\frac{\alpha V_{0}}{H^{4}}, n=0;
 \end{equation}
\begin{equation}
a^{2}=\frac{nV_{0}}{H^{4}}(\frac{H}{M})^{n}, n\neq 0.
 \end{equation}
 When $n=0$, we have $a^{2}=3\alpha m_{pl}^{2}/H^{2}$. If $H< m_{pl}$, $a^{2}> 3\alpha $  which means $a$ is larger than $1$.
 When $n\geq 4$, $V_{0}/(M^{n}m_{pl}^{4-n})<1$ yields
$a^{2}<n(H/m_{pl})^{n-4}$, and as a result when $H< m_{pl}$, $a$ is
smaller than $1$. When $n=2$, we just have
$a^{2}=6m_{pl}^{2}/M^{2}$, which mean $a$ can not be decided.

 Because the effective regions of the  dissipative parameter depend on the
temperature, it is important  to investigate how $\phi/T$ evolves:
\begin{equation}
\label{hphiT}
\frac{\phi}{T}=\frac{\phi_{i}}{T_{i}}(\frac{\phi}{\phi_{i}})^{(7-2n)/(4+m)}
=[4C_{r}C_{\phi}a^{-4}(\frac{\phi}{H})^{7-2n}]^{1/(4+m)}.
 \end{equation}
As $\phi$  deceases, $\phi/\phi_{i}<1$, the evolution of $\phi/T$
depends on the value of $n$, and the value of $\phi/T$ depends on
$a$ and $\phi/H$ remarkably.

\subsection{Slow roll parameters in hybrid-like potential}
\label{sec5a}

As in monomial potential, we first  pay  attention to the slow-roll
parameters. In hybrid-like potential,
\begin{equation} \label{hepsilon}
\epsilon\approx\frac{n^{2}}{2}\frac{\phi^{2(n-1)}m_{pl}^{2}}{M^{2n}}=\frac{n\eta}{2(n-1)}(\frac{\phi}{M})^{n}.
 \end{equation}
  As
$\phi\ll M$, we have $\epsilon<\eta$, and the slow-roll requirement
for $\epsilon$ can be easily satisfied as long as $|\eta/r|<1$.

 The parameter $\beta$  can be written in the form of
$\epsilon$ and $\eta$:
\begin{equation}
 \beta=\frac{2m}{4+m}\eta-\frac{m}{4+m}\epsilon-\frac{4(m-1)}{4+m}\frac{\eta}{n-1}.
 \end{equation}
 If $\epsilon$ and $\eta$ could
 satisfy
 the slow-roll conditions, so could $\beta$.
 Combined with the discussions before,  $|\eta/r|<1$ is the key
 relation to the slow-roll requirements.

Due to the hybrid-like potential,  the parameter $
r=[C_{\phi}^{4}(a^{4}/4C_{r})^{m}(\phi/H)^{2nm-6m+4}]^{1/(4+m)}$,
thus
\begin{equation}\label{eqh1}
|\frac{\eta}{r}|=|n-1|[(4C_{r})^{m}a^{-2m+8}C_{\phi}^{-4}(\frac{\phi}{H})^{-mn+4n+4m-12}]^{1/(4+m)}.
 \end{equation}

Furthermore, the condition $|\eta/r|<1$ can explain why during
inflation the potential always  takes the dominant role, if we
rewrite Eq. (\ref{eqh1}) as
$\rho_{r}/V=12^{-1}(n-1)^{-1}(\eta/r)(\phi/M)^{n}$. As $\phi\ll M$,
the condition $|\eta/r|<1$
 results in $\rho_{r}/V<1$.

 The slow-roll parameter $c$  will give a strong constraint on the
model, which is
\begin{equation}
\label{cc}
 |c|=|m+\frac{(1-m)(4+m)}{m-3+2n}|<4.
 \end{equation}
In the $m=-1,0,1,3$ cases, the exact value of $n$ satisfying Eq.
(\ref{cc}) can be seen in Table \ref{tab3}. If $m=1$, this
constraint is satisfied by all values of  $n$. But in the $m=-1,0,3$
cases, not every $n$ is suitable, e.g if $m=-1,0$, the standard case
$n=2$ is ruled out; if $m=3$, the $n=0$ case is excluded as well.


 \begin{table}[ht]
\caption{The suitable range of $n$ for $c$, and the value of
$n_{s}-1$, and the suitable $n$ for the red spectrum and the blue
spectrum in hybrid-like potential.}
\begin{center}
  \begin{tabular}{l@{\hspace{2mm}}l@{\hspace{2mm}}l@{\hspace{2mm}}l@{\hspace{2mm}}l@{\hspace{2mm}}l@{\hspace{2mm}}cc}

    \hline
&$m$&$|c|<4$&$n_{s}-1$&red spectrum&blue spectrum \\
\hline
 &$-1$&$n>13/5,$
 $n<1$&$\frac{-3\eta}{r}[\frac{3-n}{n-1}+\frac{n}{2(n-1)}\frac{\phi^{n}}{M^{n}}]$&$1<n<3$&$n<1, n>3$\\
 &$0$&$n>2,$ $n<1$&$\frac{-3\eta}{r}[\frac{-2n+5}{4(n-1)}+\frac{3}{8}\frac{n}{n-1}\frac{\phi^{n}}{M^{n}}]$&$1<n<5/2$&$n<1,n>5/2$\\
 &$1$&any n&$\frac{-3\eta}{r}(\frac{-1}{5}+\frac{3}{10}\frac{n}{n-1}\frac{\phi^{n}}{M^{n}})$&none&all n \\
 &$3$&$n>1,$  $n<-7$& $\frac{-3\eta}{r}[\frac{n-7}{7(n-1)}+\frac{3}{14}\frac{n}{n-1}\frac{\phi^{n}}{M^{n}}]$&$n<1,n>7$&$1<n<7$\\
\hline
  \hline
  \end{tabular}
\end{center}\label{tab3}
\end{table}

\subsection{The power spectrum in hybrid-like potential}
\label{sec5b}
 We now turn to the density perturbations which are caused by the
thermal fluctuations. To calculate the magnitude of the thermal
fluctuations, we present the definite expressions of $T/H$ and
$P_{R}^{1/2}$ below:
\begin{equation} \label{eqh2}
\frac{T}{H}=[(4C_{\phi}C_{r})^{-1}a^{4}(\frac{\phi}{H})^{m-3+2n}]^{1/(4+m)},
 \end{equation}
\begin{equation}  \label{eqh3}
P_{R}^{1/2}=3^{-1/4}[(4C_{r})^{-5m-2}C_{\phi}^{18}a^{4(3m-6)}(\frac{\phi}{H})^{6(mn-2n-4m+5)}]^{1/(16+4m)}.
 \end{equation}

In Eqs. (\ref{eqh2}) and (\ref{eqh3}), a small $C_{\phi}$, which is
favored by the observations, will raise the value of $T/H$, and
suppress the amplitude of the power spectrum simultaneously.
Moreover, in Eqs. (\ref{eqh2}) and  (\ref{eqh3}) the explicit
dependence on the value of $a$ and $\phi/H $ should also be noted.


 \begin{table}[ht]
\caption{The value of $dn_{s}/d\ lnk$, the proper range of $n$ for
the negative running, positive running of the spectrum in the
$m=-1,0,1,3$ cases in hybrid-like potential.}
\begin{center}
  \begin{tabular}{l@{\hspace{2mm}}l@{\hspace{2mm}}l@{\hspace{2mm}}l@{\hspace{2mm}}l@{\hspace{2mm}}l@{\hspace{2mm}}cc}

    \hline
&$m$&$dn_{s}/d\ lnk$&negative running&positive running  \\
\hline
 &$-1$&$\frac{2\eta^{2}}{(n-1)^{2}r^{2}}[\frac{5n-16}{3}(-3n+9+\frac{3n}{2}\frac{\phi^{n}}{M^{n}})+\frac{3n^{2}}{2}\frac{\phi^{n}}{M^{n}}]$&$n<3,n>16/5$& $3<n<16/5$\\
 &$0$&$\frac{3\eta^{2}}{2(n-1)^{2}r^{2}}[(n-3)(-2n+5+\frac{3n}{2}\frac{\phi^{n}}{M^{n}})+\frac{3n^{2}}{2}\frac{\phi^{n}}{M^{n}}]$&$n<5/2,n>3$ & $5/2<n<3$\\
 &$1$&$\frac{6\eta^{2}}{5(n-1)^{2}r^{2}}[\frac{3n-8}{5}(-n+1+\frac{3n}{2}\frac{\phi^{n}}{M^{n}})+\frac{3n^{2}}{2}\frac{\phi^{n}}{M^{n}}]$ &$n<1,n>8/3$&$2<n<8/3$\\
 &$3$&$\frac{6\eta^{2}}{7(n-1)^{2}r^{2}}[\frac{n}{7}(n-7+\frac{3n}{2}\frac{\phi^{n}}{M^{n}})+\frac{3n^{2}}{2}\frac{\phi^{n}}{M^{n}}]$&$0<n<7$& $n>7$\\
\hline
  \hline
  \end{tabular}
\end{center}\label{tab4}
\end{table}

Apply the hybrid-like potential into Eq. (\ref{index}), then we can
give the index of the spectrum
\begin{equation}\label{hns}
n_{s}-1=\frac{-3\eta}{r}(\frac{mn-2n-4m+5}{(4+m)(n-1)}+\frac{3}{2(4+m)}\frac{n}{(n-1)}\frac{\phi^{n}}{M^{n}}).
 \end{equation}
Combined Eq. (\ref{eqh1}) with Eq. (\ref{hns}), the running of the
spectrum is
 \begin{eqnarray}
 \label{hrunning}
 \nonumber
\frac{dn_{s}}{d\ln k}&=&\frac{6\eta^{2}}{r^{2}}[\frac{-mn+4n+4m-12}{(4+m)^{2}(n-1)^{2}}(-2n+mn-4m+5+\frac{3n}{2}\frac{\phi^{n}}{M^{n}})\\
& & +\frac{3n^{2}}{2(4+m)(n-1)^{2}}\frac{\phi^{n}}{M^{n}}].
 \end{eqnarray}
That $n-1=0$  is  not a real singularity for the two equations above
because from Eq. (\ref{eqh1}) we can see the $\eta/r$ expression
contains the $n-1$ term too. As $\phi\ll M$, the terms which contain
$n\phi^{n}/M^{n}$ can be ignored in Eqs. (\ref{hns}) and
(\ref{hrunning}). In Table \ref{tab3}, we present  the analytic form
of $n_{s}-1$ in the $m=-1,0,1,3$ cases,  the proper $n$ for the red
spectrum and  blue spectrum as well. In Table \ref{tab4}, we list
the proper $n$ for the negative running and also for the positive
running of the spectrum. In particular when $m=1$, the power spectra
are always blue. In the following discussion, We briefly consider
the cases $n=0,2$ specially. When $n=0$, in the $m=-1,0,1$ cases,
the spectra are blue with negative running; in the $m=3$ case, the
spectrum is red without running of the spectrum. When $n=2$, the
spectra are red with negative running in the $m=-1,0$ cases,  blue
with negative running in the $m=1$ case, blue with positive running
in the $m=3$ case. In ordinary inflation, when $n=0$, the spectrum
is red and the running is positive;  when $n=2$, the index is blue
and no running \cite{Lyth:1998xn}. In the warm inflation the index
of the spectrum and the running can be closer to the observations in
some cases \cite{Hall:2003zp}.

In the hybrid-like potential which has two parameters after $n$
fixed, the discussions on the amplitude of the density perturbations
will be more involved than those in the monomial potential. In the
$m=3$ case, the index of $a$  term in Eq. (\ref{eqh3}) is positive
while in the $m=-1,0,1$ cases the index of $a$ term is negative. For
discussional convenience, we divide the dissipative coefficients
into two classes, the $m=3$ case and the $m=-1,0,1$ cases. Further
considering the magnitude of $a$, in the $m=-1,0,1$ cases we divide
the hybrid-like potentials into three situations $n=0$, $0<n<4$ and
$n\geq4$, while in the $0<n<4$ cases we only focus on the $n=2$ case
because of the undetermined $a$.

When $m=3$, we notice that the $n=0$ case is improper for the
request from the slow-roll parameter $c$  shown as in Table
\ref{tab3}. Furthermore, the requirement $T/H>1$ means:
\begin{equation}
\label{h3a1} a>(4C_{\phi}C_{r})^{1/4}(\frac{\phi}{H})^{-n/2},
 \end{equation}
and the condition $|\eta/r|<1$ yields
\begin{equation} \label{h3a2}
a<(|n-1|)^{-7/2}(4C_{r})^{-3/2} C_{\phi}^{2}(\frac{\phi}{H})^{-n/2}.
 \end{equation}
Considering the two equations above, we obtain
\begin{equation}
  \label{h3cr} C_{\phi}>4(n-1)^{2}C_{r}.
 \end{equation}
That means though the observations favored  small $C_{\phi}$, this
model still requires $C_{\phi}$ has a lower bound which is at order
${\cal O}(C_{r})$. Applying Eq. (\ref{h3a1}) to Eq. (\ref{eqh3}),
the amplitude of the power spectrum gives a constraint as below
\begin{equation}
 \label{h3phiH}\frac{\phi}{H}>1.3
C_{r}^{1/3}C_{\phi}^{-1/2}P_{R}^{-1/3}\approx
10^{3}C_{r}^{1/3}C_{\phi}^{-1/2}.
 \end{equation}
 Through as $C_{\phi}$ increases, the lower
limit of $\phi/H$ will decrease. If $C_{\phi}$ is of order  ${\cal
O}(C_{r})$, $\phi/H$ should be larger than $10^{3}$. The large
$\phi/H$ means besides $\phi\ll M$, $\phi$  has a lower bound. Put
Eq. (\ref{h3a2}) into Eq. (\ref{hphiT}), and we can get
\begin{equation}
\frac{\phi}{T}>4(n-1)^{2}C_{r}C_{\phi}^{-1}\frac{\phi}{H}.
 \end{equation}
Assuming that the order of $C_{\phi}$ is the same order as  $C_{r}$,
and the value of $\phi/H$ is larger than $10^{3}$, $\phi/T$ could be
larger than $1$, then the model will require a low temperature
regime which could be fulfilled in the $m=3$ case. However,
considering Eqs. (\ref{h3cr}) and (\ref{h3phiH}), we conclude that
$C_{\phi}$ and $\phi/H$ should be matched to satisfy the
requirements warm inflation  for $m=3$ as Ref.
\cite{BasteroGil:2006vr} mentioned.

In the  following discussions, we will focus on the $m=-1,0,1$
cases, which are similar on the behaviors. Based on the value of
$a$, we divide the potential forms into three cases ($n=0$,
$n\geq4$, $0<n<4$), and in the $0<n<4$ regime we will focus on the
$n=2$ case particularly.

When $n=0$, $a$ is larger than $1$ as mentioned above.  And in the
$m=-1,0,1$ cases, $|\eta/r|<1$ leads to
 \begin{equation}
\label{hetaH}\frac{\phi}{H}>
[C_{\phi}(4C_{r})^{-m/4}a^{(m-4)/2}]^{1/(m-3)},
 \end{equation}
and $T/H>1$ leads to
\begin{equation}
\label{hTH}\frac{\phi}{H}<(4C_{\phi}C_{r}a^{-4})^{1/(m-3)}.
 \end{equation}
The two equations above give out the regions that $\phi/H$ must
satisfy. Considering Eqs. (\ref{eqh3}) and (\ref{hetaH}), we obtain
\begin{equation}
\label{h0ps}
P_{R}^{1/2}>3^{-1/4}(4C_{r})^{(-5m+9)/4(m-3)}[a^{3(m-1)}C_{\phi}^{-3/2}]^{1/(m-3)}.
 \end{equation}
The index of $C_{\phi}$ in the above equation is
 nonnegative when $m=-1,0,1$, so is $a$. As $a$ is larger than or equals to $1$, the terms in the square brackets will be larger
than or equal to $1$  even through $C_{\phi}$ is
 of order ${\cal O}(1)$. The other terms on the right hand of Eq.
(\ref{h0ps}) could not make both $P_{R}^{1/2}\approx 5\times
10^{-5}$ and Eq. (\ref{h0ps})  satisfied simultaneously. So when
$n=0$, the warm inflation is not workable.

Now we turn to the  $n\geq4$ cases,  where $a$ is  smaller than $1$.
The requirement $T/H>1$ gives
\begin{equation} \label{h014TH}
\frac{\phi}{H}>(4C_{\phi}C_{r}a^{-4})^{1/(m-3+2n)},
 \end{equation}
and the above equation shows  that $\phi/H$ is larger than $1$. The
condition $|\eta/r|<1$ indicates
\begin{equation}
 \label{hetar}
\frac{\phi}{H}<(|n-1|)^{-4-m}(4C_{r})^{-m}C_{\phi}^{4}a^{-8+2m}]^{1/(-mn+4n+4m-12)}.
 \end{equation}
 The above two equations can give out a constraint to $C_{\phi}$,
 \begin{equation}
 \label{cphi0}
C_{\phi}>[(|n-1|)^{(m-3+2n)/n}(4C_{r})^{(m^{2}+mn+m+4n-12)/(4n+mn)}a^{-2(m-3)/n}.
 \end{equation}

 When $n\geq 4$ we give out the constraints from the amplitude of the power spectrum
by considering Eqs. (\ref{eqh3}) and ({\ref{hetar}}) in the
$m=-1,0,1$ cases respectively,
\begin{equation}
a<3^{(5n-16)/12}(|n-1|)^{(9-3n)/2}(4C_{r})^{(n-2)/12}C_{\phi}^{(4-n)/2}P_{R}^{(5n-16)/6},
m=-1;
 \end{equation}
\begin{equation}
a<3^{(-3+n)/3}(|n-1|)^{(15-6n)/6}(4C_{r})^{(n-3)/6}C_{\phi}^{(4-n)/2}P_{R}^{(2n-6)/3},
m=0;
 \end{equation}
\begin{equation}
a<3^{(8-3n)/12}(|n-1|)^{(1-n)/2}(4C_{r})^{(3n-10)/12}C_{\phi}^{(4-n)/2}P_{R}^{(3n-8)/6},
m=1.
 \end{equation}
 As $n$ increases, the upper limit of $a$ will decrease, in which the  term containing $P_{R}$
  with a positive index is the main contribution. Take $n=4$ for
example where $a^{2}=nV_{0}/M^{4}$, then for $m=-1$ we have
$a<10^{-6}C_{r}^{1/6}$,  for $m=0$ we have $a<6\times
10^{-7}C_{r}^{1/6}$, and for $m=1$ we have
$a<3.4\times10^{-7}C_{r}^{1/6}$. These requirements mean that in
order to satisfy the low amplitude of the power spectrum, $a$ should
be very small, then $V_{0}$ should be very small compared to
$M^{4}$. As $a$ is small,  Eq. (\ref{cphi0}) shows $C_{\phi}$
is larger than a small value.
So the constraint to $C_{\phi}$ is loose.

To evaluate the suitable working temperature, put Eq. (\ref{hetar})
into Eq. (\ref{hphiT}), and we can get
\begin{equation}
\frac{\phi}{T}>(n-1)^{2n-7}(4C_r)^{(-3m+mn+4n-12)/(4+m)}C_{\phi}^{(16-4n-mn+4m)/(4+m)}a^{-2}]^{1/(-mn+4n+4m-12)}.
 \end{equation}
When $n\geq4$, in the $m=-1,0,1$ cases, assuming $C_{\phi}$  is at
order ${\cal O}(1)$, that $a<1$  will lead to $\phi/T>1$ which means
the low temperature regime is needed. Then the $m=-1,0$ cases can
work, while the $m=1$ case can not.

When $0<n<4$, as we do not know the magnitude of $a$, we only
discuss the $n=2$ case specifically. When $m=-1,0$, the requirements
from the slow-roll parameter $c$ exclude the $n=2$ case just as
shown in Table \ref{tab3}, then we only need to discuss the $m=1$
case. When $m=1$, the requirement $T/H>1$ leads to
\begin{equation}
\label{a1}
a>(4C_{\phi}C_{r})^{1/4}(\frac{\phi}{H})^{-1/2},
  \end{equation}
and  the condition $|\eta/r|<1$ gives out
\begin{equation}
\label{a2}
a<(4C_{r})^{-1/6}C_{\phi}^{2/3}(\frac{\phi}{H})^{1/3}.
 \end{equation}
These two equations above can give out $\phi/H>(4C_{r}C_{\phi}^{-1})^{1/2}$.
The amplitude of the spectrum also gives a constraint
\begin{equation}
\frac{\phi}{H}>(12^{-1}C_{r}^{-1}C_{\phi})^{1/2}P_{R}^{-1}\approx
10^{8}(C_{r}^{-1}C_{\phi})^{1/2},
 \end{equation}
which shows that we need large $\phi/H$  for a low amplitude of the
power spectrum. From Eqs. (\ref{hphiT}) and (\ref{eqh3}),
\begin{equation}
\frac{\phi}{T}>4^{1/3}(n-1)^{-2/3}(\frac{\phi}{H})^{-1/3},
 \end{equation}
where we can know that the large value of $\phi/H$ will make
$\phi/T$ smaller than $1$, and the regime of the low temperature
excludes the $m=1$ case. Noticing that the requirement from the
slow-roll parameter $c$
 excludes the $m=-1,0$ cases, the scenario does not work
well when $n=2$.

In  hybrid-like model, when $m=3$, it requires the suitable matching between $C_{\phi}$ and
$\phi/H$ to work. When $n\geq4$, it will need very small $a$, large
$\phi/H$ and large $n$ to work in the $m=-1,0$ cases. When $m=-1,0,1$,  warm
inflation can not satisfy the observations of low amplitude of power
spectrum in the $n=0$ case. When $n\geq 4$  it can not available
in  the $m=1$ case for the workable temperature regime. And it cannot
work in the $n=2$ case for the sake of parameter $c$ or the low
working temperature regime when $m=-1,0,1$.

\section{Conclusion}
\label{sec6}
 Although, the slow-roll ordinary inflation is the
dominant paradigm in the very early universe. This mechanism can be
replaced by the warm inflation in which we can not neglect the
effect from the radiation due to the interactions between the
inflaton and other fields. The warm inflation scenario can alleviate
the $\eta$ problem, and its thermal fluctuations  are responsible
for the generation of the density perturbations if whose amplitude
exceeds than that of quantum fluctuations.

In this work, based on the exact calculational results of the
dissipative coefficient in Refs.
\cite{Moss:2006gt,Yokoyama:1998ju,Berera:1998gx}, we extended the
various cases of the dissipative coefficient to a general form
$\Gamma=C_{\phi}T^{m}/\phi^{m-1}$ as a toy model. And we focused on the
strong dissipative regime where $r\gg1$. To examine the
conditions which constrain the realization of warm inflation,
firstly we gave out the slow-roll conditions and the power spectrum
in the toy model which agree with the previous results,
then we took two specific potential forms (the monomial
potential and the hybrid-like potential) to discuss the
slow-roll conditions and the power spectrum in detail. The slow-roll
conditions are different  from the ordinary inflation, which  mean
not only the number of the slow-roll parameters are increased but
also the allowed regimes of the slow-roll parameters are enlarged
which may solve can alleviatethe $\eta$ problem.
The amplitude of the power spectrum is not only depend on the Hubble
parameter and the potential, but also the temperature and the
dissipative coefficient. However, we considered
$\phi\leq m_{pl}$ to constrain the value of the inflaton field
and $C_{\phi}<10^{4}$ to constrain the number of the inflaton  fields.
based on the rationality in both theory and observation.
The field regime $\phi\leq m_{pl}$ is interested as in
the ordinary inflation the monomial potential has the causal problem.
The regime $C_{\phi}<10^{4}$ is concerned because some warm inflation
models need too large number of fields to realize. Under these strong
assumptions, in the monomial potential, the $m=0$
case is a workable model; and when $2\leq n<10/3$, the $m=1$ can
work too. In the hybrid-like potential, the $m=3$ case is workable,
and when $n\geq4$ the $m=-1,0$ cases can work as well. It seems that
the cases that $m=0,3$ are worth further research because they are
possibly  permissible in warm inflation framework. And subsequent
work on the details of warm inflation are worthy to be investigated
too. After all, it is a workable scenario which is different from
ordinary inflation.


\section*{Acknowledgements}
YZ thanks Rong-gen Cai, Hui Li, Seoktae Koh, Bin Hu  and Ya-wen Sun
for useful discussions. This work was supported in part by a grant
from Chinese Academy of Sciences, grants from NSFC with No. 10325525
and No. 90403029.

\end{document}